# Scanning tunneling microscopy study of Ni$_2$MnGa(100) surface

Jayita Nayak,[1+] Abhishek Rai,[1+] D. L. Schlagel,[2] T. A. Lograsso,[2] and S. R. Barman[1]

[1]*UGC-DAE Consortium for Scientific Research, Khandwa Road, Indore, 452001, India and*

[2]*Ames Laboratory U.S. DOE, Iowa State University, Ames, Iowa 50011-3020, USA*

[+]Both the authors have equally contributed to this work

ABSTRACT

Ni$_2$MnGa(100) surface has been investigated in the premartensite and martensite phase by using scanning tunneling microscopy. The presence of twined morphology is observed in the premartensite phase for Mn excess surface which exhibit non-equispaced parallel bands in one side of the twin boundary. Moreover, in the flat region of the surface two domains of non-periodic parallel bands corresponding to the incommensurate CDW is observed. Although, stoichiometric surface also exhibits twining but the parallel bands are equispaced and have equal corrugation. Most interestingly, coexistence of twined morphology and the CDW pattern is observed in the premartensite phase for Ni excess surface which was not reported till date. In the martensite phase for Mn excess surface, incommensurate CDW is transformed to commensurate CDW corresponding to the equispaced parallel bands. In stark contrast, stoichiometric surface exhibit parallel bands that have different periodicity in different regions. Both the voltage dependent STM and STS measurement establishes that this morphology is also related to the CDW.

# INTRODUCTION

Recently, Ni2MnGa has created immense interest due to its faster dynamic response than the conventional shape memory alloys and as well as giant magnetic field induced strain and large negative magneto-resistance which, make them suitable for device applications like actuators, sensors, magnetic refrigerator, etc. It under- goes a symmetry lowering transformation to a modulated phase with orthorhombic unit cell[1–4] when the material is cooled below the martensite transition temperature (200 K). Prior to the martensite phase, Brown et al. showed that on cooling below 260 K it transforms to pre-martensitic phase which is characterized by tripling of the repeat in one of the <110> directions occurs and persists down to the structural phase transition at TM =200 K [2]. In sprite of the fact that the basic structure is tetragonal in the martensitic phase, a weak modulation in atomic position is observed which involves periodic shuffling of atoms in the (110) plane along the [1$\bar{1}$0] direction[3, 4]. Modulation occurs due to a TA2 soft phonon mode in the transverse acoustic branch along [110] direction as indicated by diffuse x- ray scattering studies[5]. Modulation plays an important role in martensite phase since large magneto crystalline anisotropy is observed in modulated phase leading to large magnetic field induced strain which is very useful for device applications like actuators.

A typical martensite microstructure consists of combination of different twin variants in which two adjacent twins meet at well-defined interfaces called twin planes. Each of these twins have a unique orientation defined by its c-axis and the martensite phase is essentially composed of variable volume fractions of the different twins. The observation of twins in the martensite phase and the presence of surface relief was reported by Pan and James[6]. Pons *et al.* reported the evidence of internal twin within the main twin i.e. hierarchically twinned microstructure and showed that the internal twins are not regular across the material and their thickness varies in different regions[7]. Another type twining pattern has been observed by Mahnke *et al.* in which the SEM images reveal a topography contrast on the Ni$_2$MnGa film surface arising from the twin lamellae that intersect at an angle of 90◦[8]. On the other hand, strip domain structure was reported by Chernenko et al. from MFM measurements. Their focused ion beam observations indicate a perfect in-plane laminated substructure of each grain with the averaged thickness of lamella being about 0.04 nm, each lamella being a martensite twin variant and the origin of this martensite nanostructure was assigned to the crystal texture[9]. In an interesting study, Armstrong *et al.*

reported the observation of twining in the premartensite phase in $Ni_{52}Mn_{23}Ga_{25}$ if the sample is cooled slowly[10]. This is a surprising result since generally it is accepted that the premartensite phase is a micromodulated cubic phase[2].

On the other hand, neutron scattering experiment reveals the presence of a phason branch that indicates the existence of CDW in the martensite phase [11]. Moreover, Zheludev *et al.* found that the phonon softening occurs below the premartensite transition temperature TP = 260 K, with wave vector ($\zeta,\_\zeta,0$), where $\zeta$= 1/3[12]. In stark contrast, Kaufmann *et al.* reported that the modulation in the martensite phase of a Ni excess $Ni_2MnGa$ film is not related to the electronic origin but occurs due to the branching of the twin variants in the tetragonal structure[13]. Most recently, it has been demonstrated that the modulation in the premartensite phase is incommensurate[14] and the existence of CDW was established from photoemission spectroscopy by the analysis of spectral shape and the presence of pseudogap at $E_F$ [15]. Most of the reports on Ni-Mn-Ga alloys till date mainly focus on studies of microscopic, structural and physical properties in the martensite phase in larger length scales (in μm range) and involves mostly polycrystals or thin films, but the surface studies on atomic scale are rare. In this letter we report scanning tunneling microscopy (STM) study of $Ni_2MnGa$(100) surface in the premartensite and martensite phases that not only provides detailed information about the twinning pattern in nanometer scale but also afford crucial understanding about the possible origin of CDW, as reported previously by photoemission spectroscopy[15] and related studies. In addition to that, scanning tunneling spectroscopy (STS) has been used to study the electronic structure around the Fermi level ($E_F$).

## EXPERIMENTAL DETAILS

STM experiments on Ni2MnGa(100) single crystal were carried out with VT-STM (variable temperature) work station from Omicron GmbH, Germany at a base pressure of about $2\times10^{-11}$ mbar. The STM images were recorded in the constant current mode by keeping the specimen at ground potential. Electrochemically etched polycrystalline tungsten tips were used for the measurements. Image analysis have been performed by using scanning probe image processor (SPIP 6.4.1) from Image Metrology.

Ni2MnGa(100) crystal was grown by Bridgman technique and was oriented in the austenite phase by Laue back reflection and the polishing was done mechanically using quarter micron diamond paste followed by electro polishing in nitric acid and methanol[16]. Ni2MnGa(100) crystal was cleaned in-situ by repeated cycles of 0.5 keV Ar ion sputtering for 15 min followed by annealing of about 30 min. Since the surface composition of Ni2MnGa strongly depend on the annealing temperature, so the stiochiometric surface (Mn:Ni=0.55) was prepared by annealing at 720 K where as Mn excess surface (Mn:Ni=0.68) was generated by high temperature annealing at 750K[17, 18]. Sharp (1×1) low energy electron diffraction (LEED) spots of four-fold symmetry were observed at room temperature for both stoichiometric and Mn excess surface corresponding to the cubic L2$_1$ structure and the surface composition as well as the absence of surface contamination due to oxygen or carbon was confirmed by Auger electron spectroscopy. Since, the premartensite and martensite phase transitions occur at 270K and 200K respectively for the stoichiometric Ni$_2$MnGa surface, so the STM measurements were performed at 230 K and 90 K corresponding to the premartensite and martensite phases. Moreover, we are able to enhance the premartensite transition temperature(270K for the stoichiometric Ni2MnGa surface) to room temperature by simultaneous sputtering and annealing at 620 K.

## RESULTS AND DISCUSSIONS

### Premartensite phase: Mn excess surface

Interestingly, STM measurement provides the evidence of twinned morphology in the premartensite phase for the Mn excess surface, where surface corrugation of the order of 0.7±0.2 nm in the out of plane direction is observed with a twinning angle of 175±1.5∘ (Fig. 1(a-c)). The average width of the twin bands is 12±2 nm, and the presence of the twin boundaries are clearly observed in the 3D view (Fig. 1 (d). Steps of atomic height cross the twin boundaries without affecting the twin morphology (white arrows in Fig. 1(a)). A close inspection of the STM image reveals the presence of parallel bands at one side of the twin boundary and lie perpendicular to the twin boundary. Such a region is shown in an expanded scale in Fig. 1(e). The line profile (Fig. 1(f)) shows that the width of these bands (1.3±0.3 nm) as well as their corrugation (0.03±0.01 nm) are an order of magnitude less than the twins. These bands have nearly equal spacing, but in some regions the bands seem to overlap forming a pair (shown by red arrow), where the height of the second band is about half of the first.

In the premartensite phase, a different type of surface structure is observed in the flat regions of the surface that do not show the surface corrugation (Fig. 2(a)). In this case, parallel bands are observed on both sides of the domain boundary (shown by black dashed line). Interestingly, at the junction point on the boundary, they meet in a unique fashion (shown by black circle) where, two or three bands merge with one band from the other region (Fig. 2(b)). Clearly, mirror symmetry, that is characteristic of a twin boundary [19], is absent in this case which establishes that these parallel bands are not originating from the twining. Previously, similar type of CDW domains were observed in STM by Braun et al. in case of Cr(110) surface and also in 1T-TaS2 sample by Wu et al. [20, 21]. This indicates that these parallel bands are possibly related to the CDW. Moreover, at the boundary, the parallel bands from two domains meet each other at an angle 92±0.3° and the bands occur either parallel or perpendicular to the step edges, and are not influenced by the presence of step edges. From the line profile of Fig. 2(a), the parallel bands are non-equispaced (Fig. 2(b)). Moreover, these bands are of unequal heights and the maximum height variation (0.04 nm) is much smaller than the step height (0.29 nm). Clustering of bands is apparent along the blue line where three parallel bands are clustered, but later it converts into pairing of two parallel bands. On the other hand, along the red line, mostly pairing is observed. In fact, the non-periodicity of parallel bands is firmly confirmed from the FFT image (Fig. 2(c)) which exhibit several intense spots appearing along the diagonal. Enhanced noise in the central part is an artifact. Moreover, the spots along both 1-5 and a − e direction are not equispaced i.e for instance the spot-separation between 1 and 2 are closer compared to the other spots. Similar type of FFT pattern with multiple spots at unequal separation has been observed by Kim et al. for CeTe3 and by Fang *et a.l.* for TbTe3 [22, 23]. Fig. 2(e) represents a line cut in Fig. 2(d) showing five peaks at 1.115 nm$^{-1}$ (=0.07 b*), 1.53 nm−1 (=0.1 b*), 2.23 nm−1 (=0.15 b*), 3.35 nm$^{-1}$ (= 0.224b*) and 4.53 nm$^{-1}$ (=0.3 b*) respectively, where b* represents the reciprocal lattice vector along b direction. Similar spot profile was also observed along a – e direction. Moreover, from the spot profile analysis of FFT images both the authors identified the CDW wave vectors and assigned these features to be the signature of incommensurate CDW. Interestingly, we found that the parallel bands survives even at higher bias voltages (1.7V) (Fig. 2(f)) but the CDW related feature changes, which is reflected in the FFT image (Fig. 2(g)) where only four spots are observed instead of five in contrast to FFT image obtained at 0.6 eV bias voltage (Fig. 2(d)). It is important to mention that the spot close to the centre is an artifact. The changes of FFT pattern with the change of bias voltage provides an evidence of CDW that has been established by both experimental and theoretical studies [23, 24]. From all these observations we can conclude that the occurrence of of parallel bands represents the existence of CDW. Furthermore, it establishes that the origin of CDW is related to the electronic structure not due to twining in stark contrast to the adaptive martensitensite model proposed by Kaufmann *et al*.[13].

Fig. 3(a) and (b) presents an atomic resolution image of the parallel bands in the premartensite phase of $Ni_2MnGa$. From the line profile analysis (Fig. 3 (c)), although no in-plane displacement of atoms has been observed but it is evident that all the bands does not contain equivalent no of atoms i.e. for example lower corrugation bands can accommodate nearly 3-4 atoms whereas higher corrugation bands contain 6-7 atoms. Now, the FFT image (Fig. 3(d)) exhibits two distinct features-(a) sharp diffraction pattern (black circles) and the satellite spots (shown by arrows) resulting from the atomic resolution and (b) the diagonal line consisting of many bright spots at the center corresponding to the parallel bands due to CDW. Surprisingly, the lattice parameter obtained in the premartensite phase is 0.46±0.01 nm which is significantly higher compared to it's parent austenite lattice parameter (0.41 ±0.02 nm) and the angle between the lattice vector is about 93±0.5 ° which exhibit the change of crystal structure in the premartensite phase from cubic (austenite) to monoclinic. Since the CDW transition occurs along the b direction so the satellite spots observed along with the main diffraction spots and it provides important information about the CDW wave vector. The main satellite spot is shown by the white arrow whereas the red arrow represents the diffuse satellite spots. From the line profile (Fig. 3(f)) along A-C direction the satellite spots appear at B=3.48 $nm^{-1}$ and C=4.76 $nm^{-1}$ from main spot (A) whereas along a-c the spot separation are 2.93 $nm^{-1}$ and 3.58 $nm^{-1}$. Surprisingly, the main satellite spot separation (3.58 nm−1) is closely related to the spot profile 4 (3.35 nm−1) in Fig. 2(e). This establishes that CDW wave vector is $q_{CDW}$=3.58 $nm^{-1}$= 0.224b* and represented by spot 4 in Fig. 2(d). From XRD measurement Singh et al. reported CDW wave vector to be $q_{CDW}$=0.33761 c* for stoichiometric $Ni_2MnGa$(100) surface in the premartensite phase[14]. The reason for achieving lower value of $q_{CDW}$ from STM measurement is possibly related to the fact that the surface was Mn excess in contrast to the stoichiometric sample studied by Singh et al. Another possible reason would be that STM provides local information about the surface only where as XRD gives average information of the bulk which may be considerably different from the surface. Similar discrepancy between STM and XRD was reported for another CDW system TbTe3 which was ascribed to the enhanced surface sensitivity for the STM [23]. Now if we consider the central part of the FFT image (Fig. 3(e)) corresponding to the parallel bands, we find that only three spots are apparent and their spot separation is similar to that of spot 3, 4, and 5 in Fig. 2(e) which further confirms the validity of the CDW wave vector.

## Premartensite phase: stoichiometric surface

Twined microstructures are also present in the premartensite phase (230K) for the stiochiometric surface (Fig. 4(a)) but in some region of the surface the twin boundary is horizontal to the image plane (shown by the black rectangle) where as in some other region it is in vertical direction (shown by the white rectangle). The presence of parallel bands are also observed in one side of the twin boundary (Fig. 4(b)). Interestingly, twin bands on both side of the twin boundary are quite symmetric (17±1 nm) (Fig. 4(c)). The average surface corrugation of the twin bands is 0.5±0.2 nm (in the out of plane direction) and the twining angle is 176±0.8∘. In stark contrast to the Mn excess surface, zoomed image (white rectangle) of the parallel bands (Fig. 4(d)) demonstrate the presence of equispaced parallel bands instead of pairing which is a charecteristics of Mn excess surface. The corrugation of the band is 0.03±0.01 nm and the width is 2±0.1 nm (Fig. 4(e)). Thus, the bands are much wider compared to the Mn excess surface (0.03±0.01 nm). No significant changes of band features have been detected with the change of bias voltage (Fig. 4(f)) indicates that they are purely topographic in origin and not related to the CDW. We are succeeded to enhance the premartensite transition temperature (260K) of $Ni_2MnGa$ to room temperature (300K) by hot sputtering which is confirmed by the LEED pattern (Fig. 5(a)). We have performed hot sputtering by keeping the substrate to subsequently lower temperature (620K) than the of stoichiometric surface (720K). D'Souza et al. reported that with the decrease of annealing temperature the surface becomes Ni excess and as a result premartensite transition temperature is enhanced[17]. This explains the occurrence of premartensite phase at room temperature. Surprisingly, we find the coexistence of twined morphology (white rectangle) and the CDW ordering (black rectangle) which was not observed in stoichiometric as well as Mn excess surface (Fig. 5(b)). For the twining, parallel bands are observed but the separation between the bands varies randomly from 1.6 to 2.4 nm but no pairing is observed unlike to the Mn excess surface. Most interestingly, CDW pattern (black rectangle) is superimposed on the corrugated surface resulting from the twining (Fig. 5(c)) in stark contrast to both Stoichiometric and Mn excess surface where CDWpattern was observed in the flat region of the surface (Fig. 2(a)). The contribution of twined structure has been removed by using filtering method for the better clarity of presentation and to extract more information about the CDW structure (Fig. 5(d)). In the FFT image of (Fig. 5(d)) no sharp diffraction spot has been identified corresponding to the parallel bands (Fig. 5(e)), which indicate that the bands are not periodic which is also con- firmed by the line profile (Fig. 5(f)) analysis. From the line profile it is clear that the width of the bands varies between 1.4 nm to 2.2 nm in a random fashion and the angle between the bands originating from two domains are $113\pm 1°$. This is a very important result since this type CDW with irregular spacing is unique, which has not been reported so far.

## Martensite phase for Mn excess surface

It is surprising that the STM topographic image of $Ni_2MnGa(100)$ in the martensite phase at 80 K also exhibits parallel bands (Fig. 6(a)). However, there is a basic difference of these parallel bands and those observed in the premartensite phase: here the parallel bands are equispaced ($1.5\pm0.08$ nm) (Fig. 6(b)) as observed in the FFT image (shown by white circles). Similar parallel bands of spacing 4.2 nm was previously observed by Braun *et al.* on Cr(110) and the authors assigned this feature to be the evidence of surface CDW[21]. In Cr/W(110), Hsu *et al.* reported similar type of parallel bands that also originated from CDW and found the variation of CDW wave- length with the change of Cr coverage[25]. Thus the presence of equispaced parallel bands represents the existence of commmensurate CDW. Time dependent STM image of the same surface reveals the evolution of nano-twins (which appear as bright patches) after 65 mins. of scanning. (Fig. 6(c)). These nano twins are irregular in shape and have an average height of about $0.13\pm0.02$ nm, as ob-served from the histogram. From particle size analysis, the average area of these twins is about 3 $nm^2$. In the martensite phase, presence of similar stripe-like pattern, albeit of order of magnitude larger size (width 40 nm) was observed by Chernenko et al. from the focused ion beam measurements on $Ni_{53.5}Mn_{23.8}Ga_{22.7}$ thin films. They concluded that these correspond to the twin variants in the martensite phase[9]. From the FFT image (Fig. 6(d)) no quantitative change in

CDW pattern has been observed due to the evolution of namo-twins. From the atomic resolution image (Fig. 6(e)) we found that the lattice parameter is 0.44 nm which is increased compared to the austenite lattice parameter (0.41 nm). But the most interesting observation is that the angle between the inplane lattice vector is 100° which has not been observed so far by any other measurement.

## Martensite phase for stoichiometric surface

Stoichiometric $Ni_2MnGa$ surface also exhibits the presence of parallel bands (Fig. 7(a)) but surprisingly no sharp diffraction spots have been observed in the FFT image (Fig. 7(b)) representing that the bands are not periodic. After removing the features related to the steps it is clearly evident that different regions (shown by coloured lines) exhibit different periodicity i.e. the periodicity in red, blue and green regions are 1.59, 1.57 and 1.67 nm respectively. In stark contrast to this, when -0.1V bias voltage is applied a strong suppression of intensity of the bands are clearly observed. Although, the authors found that parallel bands persist even at a gap voltage of 1 V. This indicate that the features are not related to the surface topography and have electronic origin. However, if we compare the voltage dependent STM images with the theoretically calculated DOS [26] as well as the STS measurement (Fig. 7(e) in the martensite phase, (Fig. 7(e)) higher DOS at 1.6 eV energy in the valence band results in a clear observation of parallel bands. On the otherhand, at -0.1 eV energy in the conduction band significantly lower DOS is attributed to a strong suppression of CDW related features. Thus, the present results indicate the occurrence of a CDW state in the martensite phases, in agreement with our earlier photoemission studies that demonstrated the existance of CDW from the detailed analysis of the shape of the spectral function that revealed the presence of pseudogap at $E_F$ [15]. Moreover, the appearance of pseudogap is further confirmed by the STS measurement (Fig. 7(e)). We have identified a feature (shown by black arrow) related to the increment of DOS at about 0.9 V bias voltage. Similar features ware observed both in the premartensite and martensite phase from the previous UPS measurement which was ascribed to the spectral wave transfer due to CDW. Although of the CDW gap is difficult to elucidate because the variations in the dI /dV at the edges are subtle. To illustrate how we chose the locations of the arrows, we zoom in on the STS spectra near the Fermi level (Fig. 7(f)) having a small change of slop in the spectra leading to an estimated width of 560 meV.

The magnitude and shape of the dI /dV curve is in close agreement to other CDW systems TbTe3 and CeTe3 reported by Fang *et al.* and Tomic *et al*. respectively[23, 27].

## CONCLUSIONS

In summary, we find the evidence of twinned morphology for both stoichiometric and Mn excess surface, although twinning is not expected in the premartensite phase. One side of the twin boundary parallel bands that are perpendicular to the twin boundary are observed. For Mn excess surface these parallel bands are non-equispaced and not of equal corrugation whereas for the stoichiometric surface these bands are equispaced and exhibit equal corrugation. Interestingly, on flat regions of the Mn excess surface that do not show the above-mentioned twins, another type of parallel bands is observed where two domains of the parallel bands meet at an angle of about $92\pm0.3°$ on the surface, but at the boundary no mirror symmetry is observed. These parallel bands are non-periodic and the analysis of the FFT image reveals that these bands are related to the CDW. Surprisingly, we find the coexistence of CDW and ntwined morphology in the premartensite phase occurred at room temperature for the Ni excess surface which was prepared by the hot sputtering. On the other hand, Mn excess surface exhibit equispaced parallel bands having separation of $1.5\pm0.08$ nm which corresponds to the commensurate CDW in the martensite phase. In stark contrast, stoichiometric surface exhibit parallel bands that possess different regions having different periodicity which is further related to the CDW as confirmed by the voltage dependent STM measurement. Moreover, STS measurement reveals the presence of pseudogap which is also a signature of CDW.

REFERENCES

[1] S. Singh, V. Petricek, P. Rajput, A. H. Hill, E. Suard, S. R. Barman, and D. Pandey, Phys. Rev. B 90, 014109 (2014).

[2] P. J. Brown, J. Crangle, T. Kanomata, M. Matsumoto, K. -U. Neumann, B. Ouladdiaf and K. R. A. Ziebeck, J. Phys. Condens. Matter 14, 10159 (2002).

[3] P. J. Webster, K. R. A. Ziebeck, S. L. Town, and M. S. Peak, Phil. Mag. B 49, 295 (1984).

[4] V. V. Martynov and V. V. Kokorin, J. Phys. III 2, 739 (1992).

[5] G. Fritsch, V. V. Kokorin, and A. Kempf, J. Phys. Condens. Matter 6, 107 (1994).

[6] Q. Pan and R. D. James, J. Appl. Phys. 87, 4702 (2000).

[7] J. Pons, V. A. Chernenko, R. Santamarta and E. Cesari, Acta Mater. 48 3027 (2000).

[8] G. J. Mahnke, M. Seibt, and S. G. Mayr, Phys. Rev. B 78, 012101 (2008).

[9] V. A. Chernenko, R. L. Anton, M. Kohl, M. Ohtsuka, I. Orue and J. M. Barandiaran, J. Phys. Condens. Matter 17, 5215 (2005).

[10] J. N. Armstrong, J. D. Felske, and H. D. Chopra, Phys. Rev. B 81, 174405 (2010).

[11] S. M. Shapiro, P. Vorderwisch, K. Habicht, K. Hradil, and H. Schneider, Europhys. Lett. 77, 56004 (2007).

[12] A. Zheludev, S. M. Shapiro, P. Wochner, A. Schwartz, M. Wall, and L. E. Tanner, Phys. Rev. B 51, 11310 (1995); A. Zheludev, S. M. Shapiro, P. Wochner, and L. E. Tanner, Phys. Rev. B 54, 15045 (1996).

[13] S. Kaufmann, U. K. Roler, O. Heczko, M. Wuttig, J. Buschbeck, L. Schultz, and S. Fahler, Phys. Rev. Lett. 104, 145702 (2010).

[14] S. Singh, J. Nayak, A. Rai, P. Rajput, A. H Hill, S. R. Barman and D. Pandey, J. Phys.: Condens. Matter 25, 212203 (2013).

[15] S. W. D'Souza, A. Rai, J. Nayak, M. Maniraj, R. S. Dhaka, S. R. Barman, D. L. Schlagel, T. A. Lograsso, and A. Chakrabarti., Phys. Rev. B 85, 085123 (2012).

[16] D. L. Schlagel, Y. L. Wu, W. Zhang, T. A. Lograsso, J. Alloys Compd. 312, 77 (2000) .

[17] S. W. D'Souza, J. Nayak, M. Maniraj, A. Rai, R. S. Dhaka, S. R. Barman, D. L. Schlagel, T. A. Lograsso, A. Chakrabarti, Surf. Sci. 606, 130 (2012).

[18] R. S. Dhaka, S. W. D'Souza, M. Maniraj, A. Chakrabarti, D. L. Schlagel, T. A. Lograsso, and S. R. Barman, Surf. Sci. 603, 1999 (2009).

[19] M. Han and F. F. Kong, J. Alloys Compd. 458, 218 (2007).

[20] Xian Liang Wu and Charls M. Lieber, Science, 243, 4899 (1989).

[21] K. -F. Braun, S. Fölsch, G. Meyer, and K. -H. Rieder, Phys. Rev. Lett. 85, 3500 (2000).

[22] H. J. Kim, C. D. Malliakas, A. T. Tomić, S. H. Tessmer, M. G. Kanatzidis, and S. J. L. Billinge, Phys. Rev. Lett. 96, 226401 (2006).

[23] A. Fang, N. Ru, I. R. Fisher, and A. Kapitulnik, Phys. Rev. Lett. 99, 046401 (2007).

[24] William Sacks, Dmitri Roditchev, and Jean Klein, Phys. Rev. B 57, 20 (1998).

[25] P. -J. Hsu, T. Mauerer, W. Wu, and M. Bode, Phys. Rev. B 87, 115437 (2013).

[26] S. W. D'Souza et al., unpublished.

[27] A. Tomic, Zs. Rak, J. P. Veazey, C. D. Malliakas, S. D. Mahanti, M. G. Kanatzidis, and S. H. Tessmer, Phys. Rev. B 79, 085422 (2009).

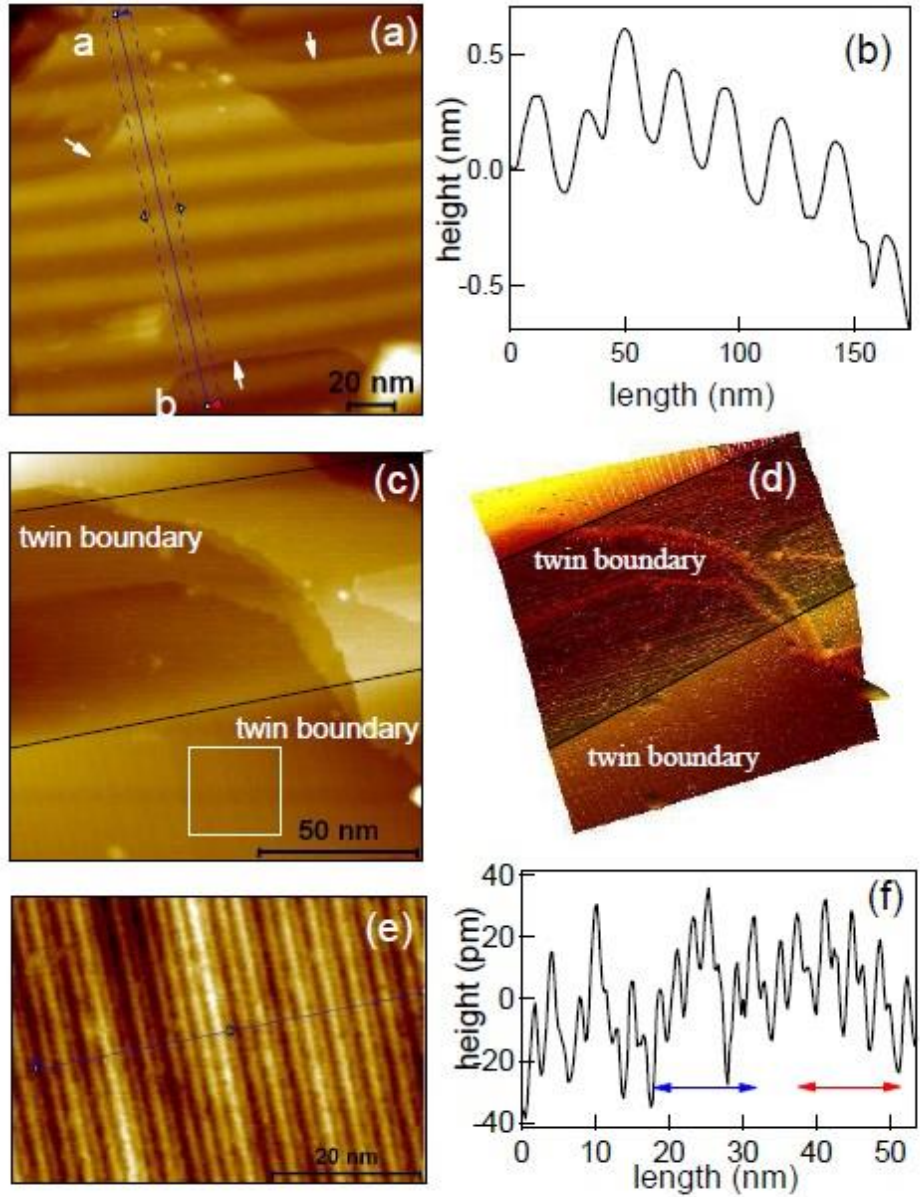

FIG. 1: STM topographic image of Ni$_2$MnGa(100) surface ($I_T$= 0.4 nA, $U_T$= 1.5 V) in the premartensite phase, (b) line profile along a-b in (a), (c) Topographic image ($I_T$= 1 nA, $U_T$= 1.3 V) of one twin indicating the presence of parallel bands in one side of the twin boundary, (d) corresponding 3D view of (c) exhibiting the presence of surface corrugation, (e) zoomed image of the region marked by a white rectangle in (c) showing parallel bands of non uniform height, (f) line profile along the blue line in (e).

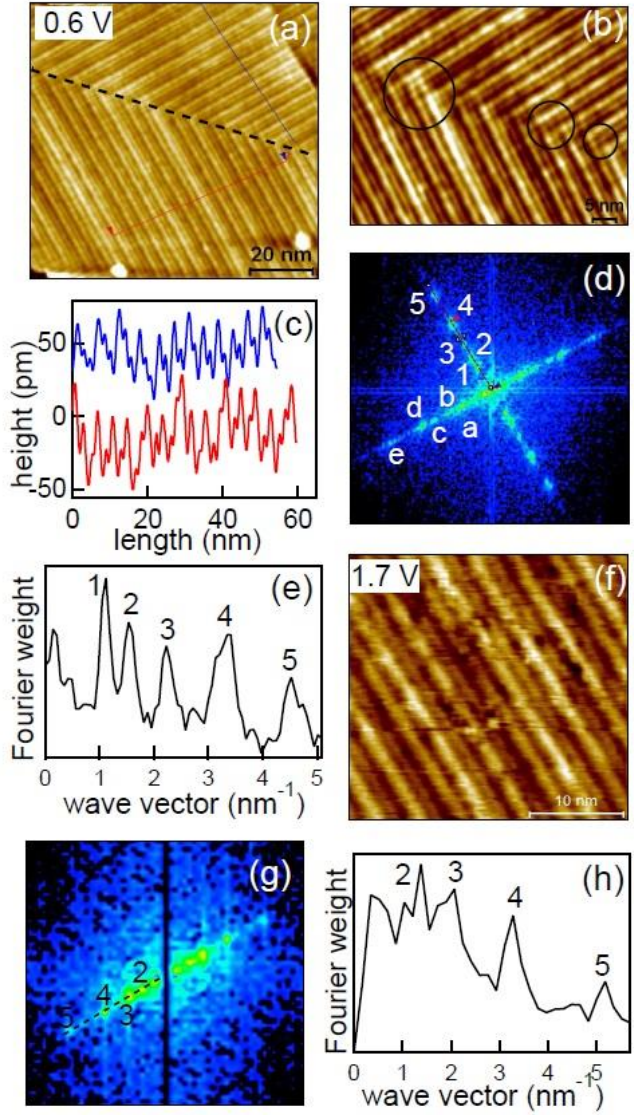

FIG. 2: (a) STM topographic image of Ni$_2$MnGa(100) surface ($I_T$= 0.5 nA, $U_T$= 0.6 V) in the premartensite phase at 230 K that reveals the presence of parallel bands, (b) zoomed image of the CDW domain boundary indicating the absence of mirror symmetry, (c) line profile corresponding to the color shown in (a), (d) corresponding FFT image of (a) showing that along both 1-5 and a − e directions the spots are non-equispaced, (e) line profile along 1-5 through the centre in the reciprocal space in (d). (f) morphology of the surface ($I_T$= 0.5 nA, $U_T$= 1.7 V) at higher bias voltage indicating the persistence of parallel bands, (g) corresponding FFT image of (f) that reveals the change of FFT pattern with bias voltage (presence of only four spot instead of five), (h) line profile along 2-5 through the centre in the reciprocal space in (g).

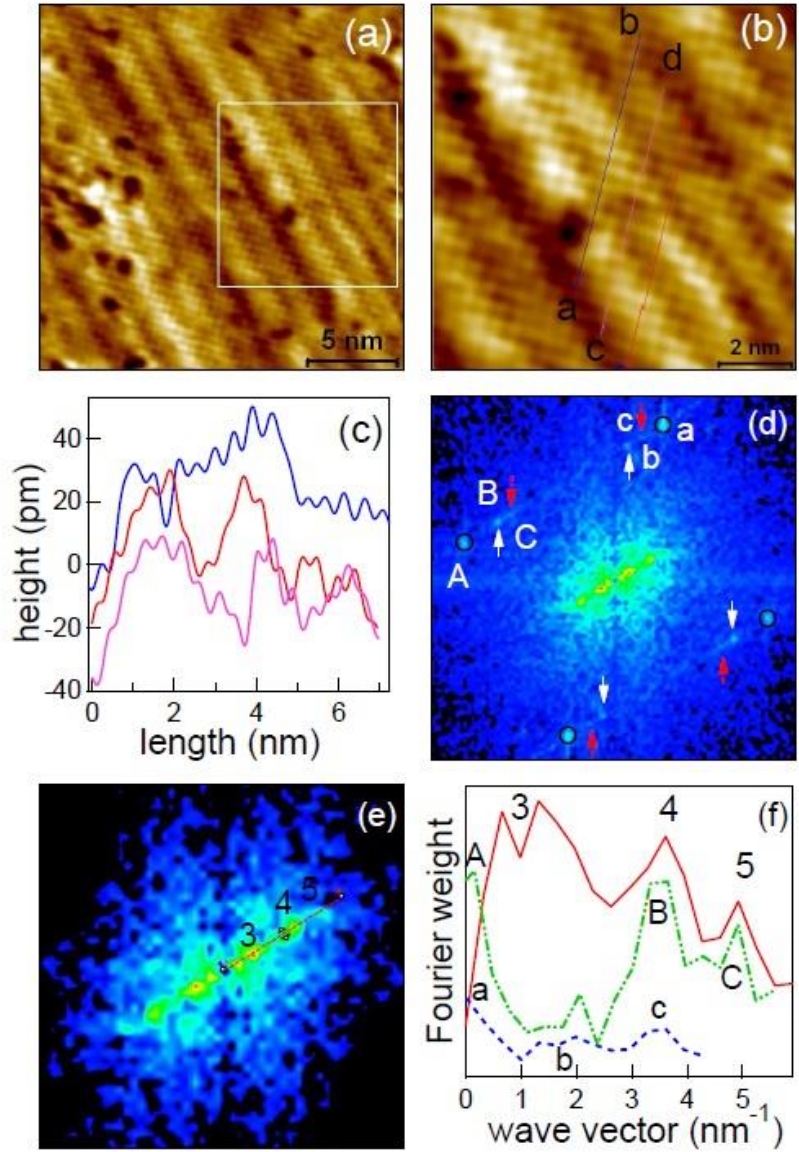

FIG. 3: (a) Atomically resolved STM image of $Ni_2MnGa(100)$ surface ($I_T$= 1 nA, $U_T$= 0.4 V) in the premartensite phase showing existence of parallel bands. (b) A zoomed image of the marked region in (a). (c) Line profile corresponding to different colors from bottom to top as shown in (b), and (d) FFT image of (a) showing diffraction spots corresponding to the atomic arrangement (black circle) along with the satellite spots (shown by the arrows) corresponding to the CDW. (e) zoomed image of the central part of FFT in (d) appears due to the presence of parallel bands. (f) line profile along a-c and A-C direction in (d) are plotted along with line profile through the centre to spot 5 in (e).

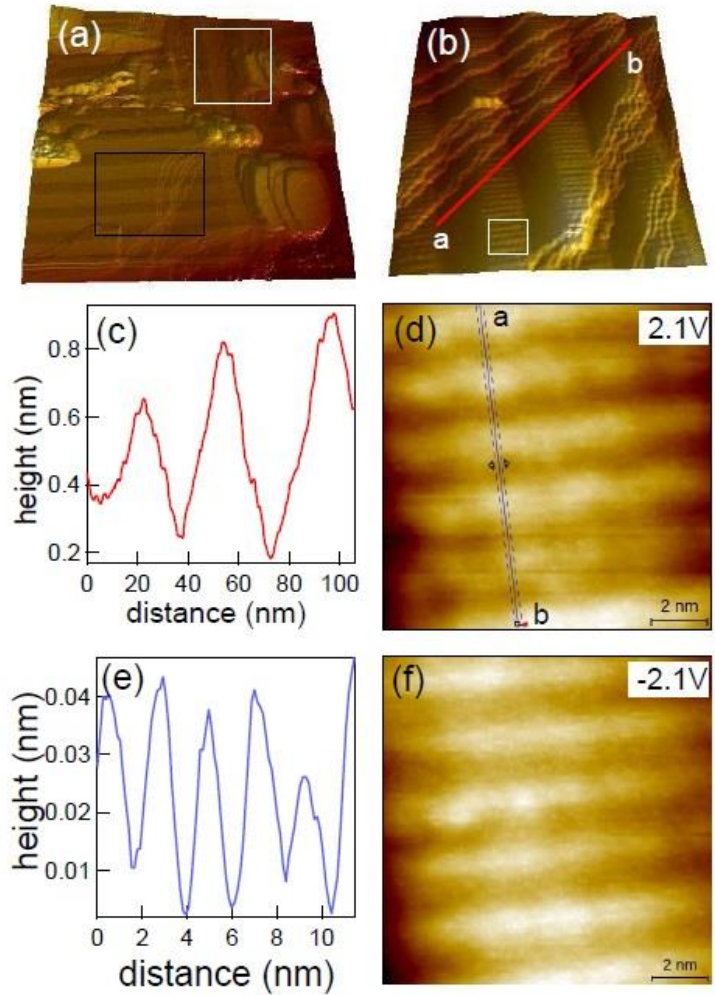

FIG. 4: 3D view of topographic STM image of $Ni_2MnGa(100)$ surface ($I_T$= 0.4 nA, $U_T$= 2.1 V) in the premartensite phase for the stoichiometric surface, (b) 3D view of the surface ($I_T$= 0.6 nA, $U_T$= 2.1 V) that reveals the presence of parallel bands in one side of the twin boundary (c) line profile along a-b in (b), (d) Zoomed image of the parallel bands marked by white rectangle in (b) (e) line profile along a-b in (d) reveals that the parallel bands are equispaced in contrast to the Mn excess surface. (f) zoomed image of the white rectangle shown in (b) with -2.1V bias voltage indicating no change of features with bias voltage.

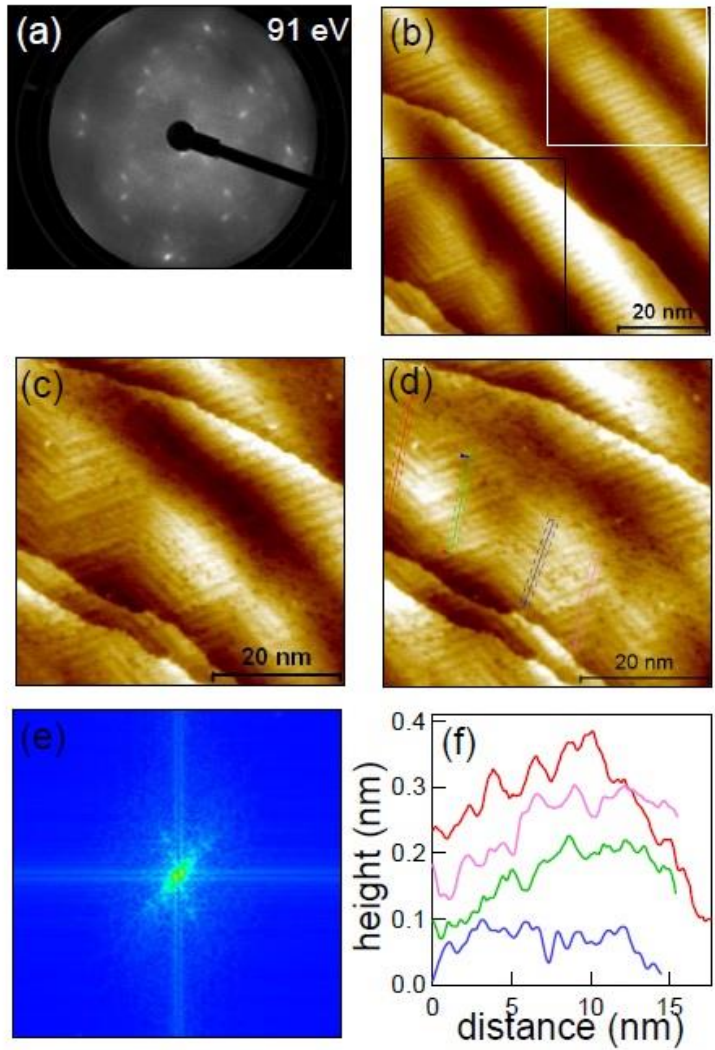

FIG. 5: (a) LEED pattern recorded with 91 eV electron beam energy for the premartensite phase of $Ni_2MnGa$ at room temperature prepared by hot sputtering. (b) STM topographic image of the surface ($I_T$= 0.3 nA, $U_T$= 1.8 V) which exhibits the coexistence of twining (white rectangle) and CDW (black rectangle). (c) Morphology of the surface ($I_T$= 0.5 nA, $U_T$= 1.9V) which contains CDW ordering superimposed with twined structure (black rectangle). (d) Flat CDW surface is extracted after removing the contribution from twining to acquire proper information about the parallel bands. (e) Corresponding FFT image of (d) indicates the lack of sharp diffraction spot due to CDW. (f) line profile corresponding to the colors shown in (d).

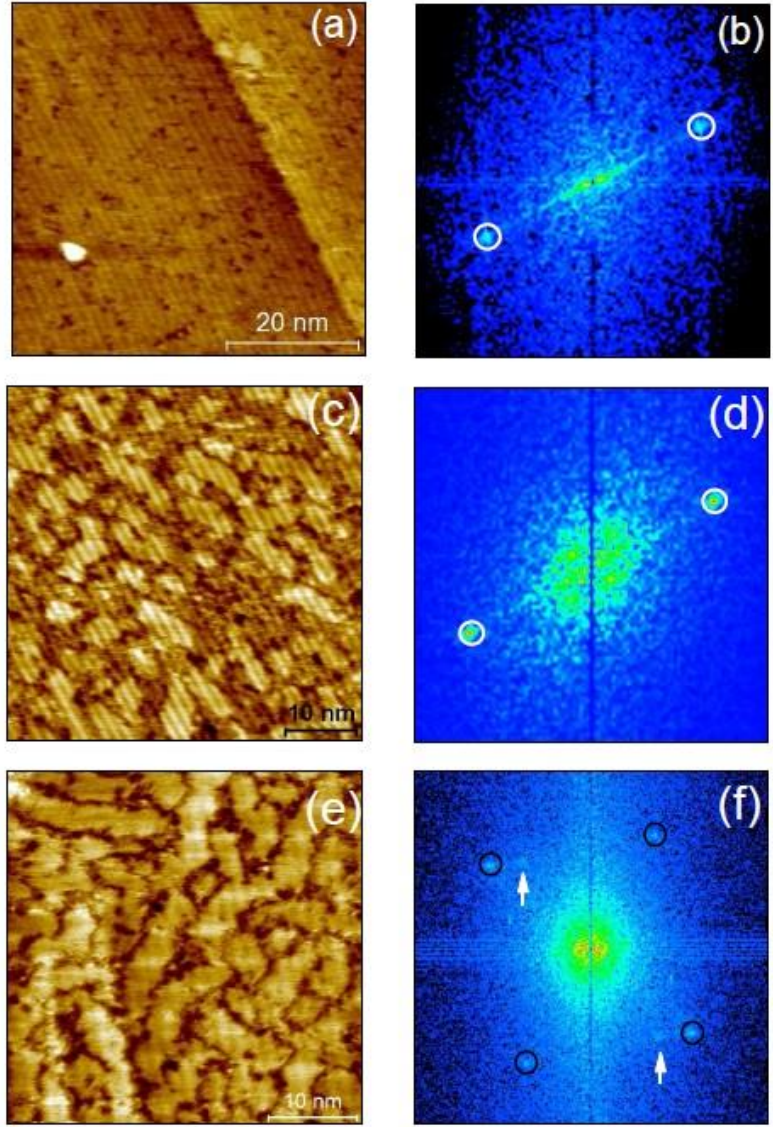

FIG. 6: (a) STM topographic image of Ni$_2$MnGa(100) surface ($I_T$= 0.2 nA, $U_T$= 1.2 V) in the martensite phase at 80 K exhibiting the presence of parallel bands, (b) FFT image of (a) exhibiting sharp diffraction spots (shown by white circle) indicating equal line spacing of the parallel bands, (c) evolution of nano-twins on the surface after 65 min($I_T$= 1.1 nA, $U_T$= -1.3 V) along with the presence of parallel bands, (d) FFT image of (c) indicating indicating no quantitative change in CDW, (e) morphology of the surface corresponding to the atomic resolution, (f) corresponding FFT image of (e).

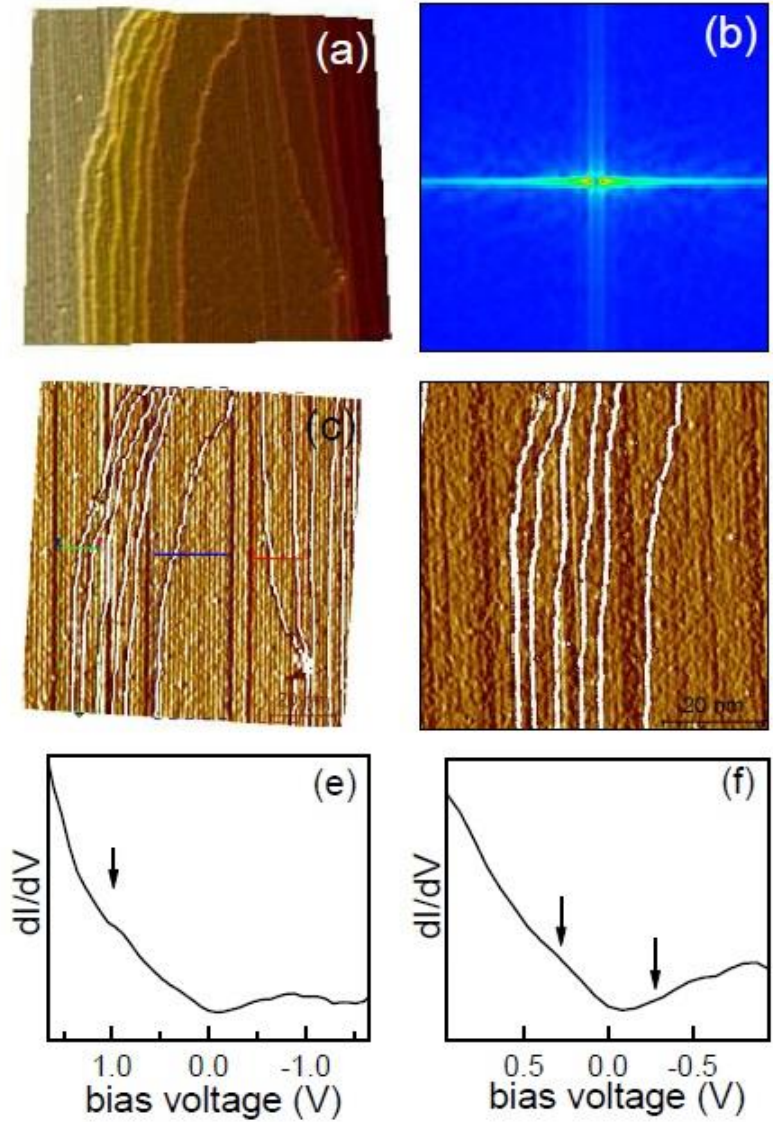

FIG. 7: (a) 3D view of STM topographic image of Ni$_2$MnGa(100) surface ($U_T$= 1.6 V, $I_T$= 0.4 nA), exhibiting the presence of parallel bands, (b) FFT image of (a) that doesn't exhibit any sharp diffraction spots corresponding to the parallel bands. (c) Detailed FFT analysis reveals that different regions (corresponding to different colours) have different band periodicity (after removing the effect of steps to enhance the week features). (d) Morphology of Ni$_2$MnGa(100) surface at -0.1 eV bias voltage $U_T$= -0.1 V, $I_T$= 0.4 nA reveals that the bands are strongly suppressed, (e) STS spectra of the same surface in the martensite phase , (f) expanded view of STS spectra near the Fermi level is shown to enhance the visibility of the subtle edges that were identified as the CDW energy gap (indicated by black arrows).